\documentclass[preprint,12pt]{elsarticle}

\usepackage{amssymb,amsmath,amsthm,bm}

\biboptions{sort&compress}

\begin{document}

\begin{frontmatter}



\title{Electromagnetic neutrino: a short review}


\author[1,2]{Alexander I. Studenikin}
\ead{studenik@srd.sinp.msu.ru}

\address[1]{Department of Theoretical Physics, Faculty of Physics, Moscow State University, Moscow 119991, Russia}
\address[2]{Joint Institute for Nuclear Research, Dubna 141980, Moscow Region, Russia}

\begin{abstract}

A short review on selected issues related to the problem of neutrino electromagnetic properties is given. After a flash look at the theoretical basis of neutrino electromagnetic form factors, constraints on neutrino magnetic moments and electric millicharge from terrestrial experiments and astrophysical observations are discussed.
We also focus on some recent studies of the problem and on perspectives.

\end{abstract}


\end{frontmatter}


\section{Introduction}

The passed two years, since the previous ICHEP conference held in Melbourne in 2012, have been celebrated by a spectacular step further in high energy physics. The prediction of the now-termed BEH symmetry breaking mechanism, attributed to  Robert Brout, Francois Englert and Peter Higgs, recently supported by an excellent job done by two CERN collaborations for discovering of the Higgs boson provides the final glorious triumph of the Standard Model.

Within the initial formulation of the Standard Model neutrinos are massless particles. However, already now it is known that the Standard Model should be extended to some more general theory in particular because of neutrinos which are the only particles exhibiting experimentally well-confirmed  properties beyond the Standard Model. This is because of  neutrino mixing and oscillations supported by the discovery of flavour conversion of neutrinos from different sources, the effect that is not possible for massless neutrinos.

In many extensions of the Standard Model, which account for neutrino masses and mixings, neutrinos acquire nontrivial electromagnetic properties that hence allow direct electromagnetic interactions of neutrinos with electromagnetic fields and charged particles or with particles which have magnetic moments. Unfortunately, in spite of reasonable efforts in studies of neutrino electromagnetic properties, up to now there is no experimental confirmation, neither from terrestrial laboratories studies nor from astrophysical observations, in favour of nonvanishing neutrino electromagnetic characteristics. However, experimentalists and theoretists are eagerly searching for them and once being experimentally confirmed they will open a window to new physics beyond the Standard Model.

The studies of neutrino electromagnetic properties has a long history. It has indeed started  long ago: the importance of neutrino electromagnetic properties was first mentioned by Wolfgang Pauli just in 1930 when he postulated the existence of the neutrino and supposed that its mass can be of the order of one of the electron. In his famous letter to ``radiative ladies and gentlemen" Pauli also discussed the possibility that neutrino  could have a magnetic moment. It is worth noting another early paper \cite{Touschek:1957} which concerns the magnetic moment of neutrino. The authors show that at the limit of mass of the neutrino equal to zero the magnetic moment also should tends to zero.

Systematic theoretical studies of neutrino electromagnetic properties have started after it was shown that in the minimally-extended Standard Model Standard with right-handed neutrinos the magnetic moment of a massive neutrino is, in general, nonvanishing and that its value is determined by the neutrino mass \cite{Fujikawa:1980yx}. For the recent reviews on the neutrino electromagnetic properties see \cite{Studenikin:2008bd, Giunti:2008ve, Broggini:2012df, Giunti:2014ixa}.

\section{Neutrino electromagnetic vertex}
\begin{figure}[h]
\begin{center}
\center{\includegraphics[width=0.4\linewidth]{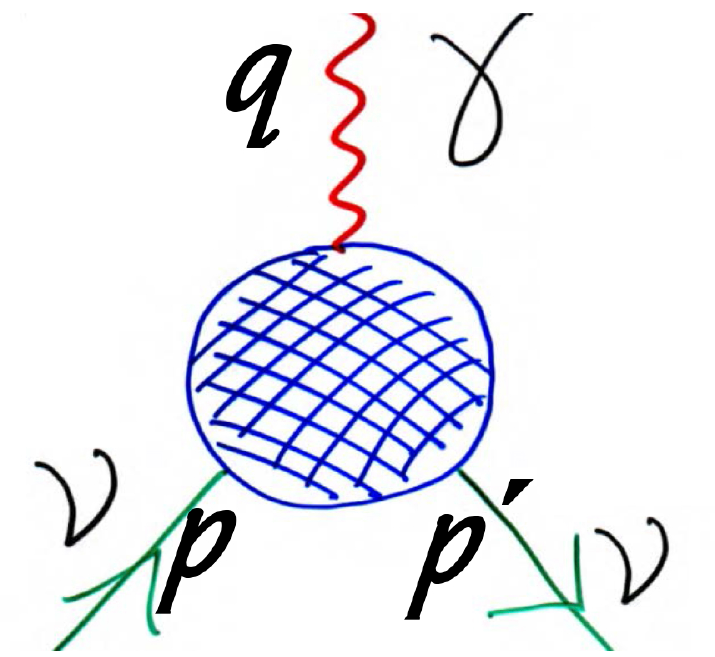}}
\caption{The effective diagramme for the electormagnetic interaction of the  initial $\psi (p)$ and final $\psi (p')$ neutrino mass states with a photon.}
\label{figure em_vertex}
\end{center}
\end{figure}

The neutrino electromagnetic properties are determined by the neutrino electromagnetic vertex
function $\Lambda_{\mu}(q)$ (see Fig.\ref{figure em_vertex}) that is related to the matrix element of the electromagnetic current between the
neutrino initial $\psi (p)$ and final $\psi (p')$ mass states
\begin{equation}\label{matr_elem}
<{\psi}(p^{\prime})|J_{\mu}^{EM}|\psi(p)>= {\bar
u}(p^{\prime})\Lambda_{\mu}(q)u(p).
\end{equation}
 The Lorentz and electromagnetic gauge invariance imply \cite{Kayser:1982br,Nieves:1981zt,
Nowakowski:2004cv,Giunti:2014ixa}
that the electromagnetic vertex function can be written in the form:
\begin{align}
\Lambda_{\mu}(q)
=
f_{Q}(q^{2}) \gamma_{\mu}
-\null & \null
f_{M}(q^{2}) i \sigma_{\mu\nu} q^{\nu} \nonumber
+
f_{E}(q^{2}) \sigma_{\mu\nu} q^{\nu} \gamma_{5}\\
+\null & \null
f_{A}(q^{2}) (q^{2} \gamma_{\mu} - q_{\mu} {q}) \gamma_{5}
,
\label{vert_func}
\end{align}
where
$f_{Q} $,
$f_{M} $,
$f_{E} $ and
$f_{A} $ are charge, dipole magnetic and electric and anapole neutrino electromagnetic form factors, and $\sigma_{\mu\nu}=\frac{i}{2}(\gamma _{\mu}\gamma_{\nu}-\gamma _{\nu}\gamma_{\mu})$.
The photon four-momentum  $q$ is given by $q=p-p'$ and form factors depend on the Lorentz invariant dynamical quantity $q^2$.

Note that the Dirac and Majorana neutrinos have quite different electromagnetic properties. This can be easily seen \cite{Giunti:2014ixa} if one accounts for constraints on form factors imposed by the the hermiticity of the neutrino electromagnetic current and its invariance under discrete symmetries transformations. In the case of Dirac neutrinos, the assumption of $CP$ invariance combined with the hermiticity of $J_{\mu}^{EM}$ implies that $f_{E}(q^2)=0$. In the case of Majorana neutrinos from the $CPT$ invariance, regardless of whether $CP$ invariance is violated or not, $f_Q(q^2)=f_M(q^2)=f_E(q^2)=0$ and only the anapole form factor $f_{A}(q^2)$ can be non-vanishing.

In general case the matrix element of the electromagnetic current (\ref{matr_elem}) can be considered between neutrino initial $\psi (p)$ and final $\psi (p')$ states of different masses $p^2 = m_i ^2\neq m_j ^2={p'}^2$. Then the vertex function and form factors are matrixes in the neutrino mass eigenstates space,

\begin{align}
\label{Lambda} \Lambda_{\mu}(q)_{ij}=\null & \null
\Big(f_{Q}(q^{2})_{ij}+f_{A}(q^{2})_{ij}\gamma_{5}\Big)
(q^{2}\gamma_{\mu}-q_{\mu}{\not
q})+\nonumber
\\ \null & \null f_{M}(q^{2})_{ij}i\sigma_{\mu\nu}q^{\nu}
+f_{E}(q^{2})_{ij}\sigma_{\mu\nu}q^{\nu}\gamma_{5}.
\end{align}
For two Dirac neutrinos in the off-diagonal case ($i\neq j$) the hermiticity by itself does not
imply restrictions on the form factors and all of them can be nonzero.
Two massive Majorana neutrinos can have either
a transition electric form factor
or a transition magnetic form factor,
but not both, and the transition electric form factor can exist only together with a transition anapole form factor, whereas the transition magnetic form factor can exist only together with a transition charge form factor.

 From the demand that the form factors at
zero momentum transfer, $q^2=0$, are elements of the scattering
matrix, it follows that  in any consistent theoretical model the form
factors in the matrix element (\ref{matr_elem}) should be gauge
independent and finite. Then, the form factors values at $q^{2}=0$
determine the static electromagnetic properties of the neutrino that
can be probed or measured in the direct interaction with external
electromagnetic fields.

\section{Neutrino electric millicharge}

It is usually believed that the
neutrino electric charge is zero. This is often thought to be
attributed to gauge invariance and anomaly cancelation constraints
imposed in the Standard Model. In the Standard Model  of $SU(2)_L
\times U(1)_Y$ electroweak interactions it is possible to get
\cite{Foot:1992ui} a general proof that neutrinos are
electrically neutral which is based on the requirement of electric
charges quantization. The direct calculations of the neutrino charge
in the Standard Model for massless (see, for instance
\cite{Bardeen:1972vi,CabralRosetti:1999ad})
and massive neutrino \cite{Dvornikov:2003js,Dvornikov:2004sj}  also prove that,
at least at the one-loop level, the neutrino electric charge is
gauge independent and vanishes. However, if the neutrino has a mass,
the statement that a neutrino electric charge is zero is not so
evident as it meets the eye. As a result, neutrinos may become
electrically millicharged particles. A brief discussion of different mechanisms for introducing millicharged particles including neutrinos can be found in \cite{Davidson:2000hf}.

The most severe experimental constraints on the electric charge of
the neutrino of the level of  $q_\nu \sim 10^{-21} e_0$ can be obtained from
neutrality of matter \cite{Bressi:2011pj}. A detailed discussion on
other constraints on $q_\nu$  can be found in \cite{Giunti:2014ixa}.

Even if the electric charge of a neutrino is vanishing, the electric
form factor $f_Q(q^2)$ can still contain nontrivial information about
neutrino static properties \cite{Giunti:2014ixa}. A neutral particle can be characterized
by a superposition of two charge distributions of opposite signs so
that the particle form factor $f_Q(q^2)$ can be non zero for
$q^2\neq 0$.  The mean charge radius (in fact, it is the charged
radius squared) of an electrically neutral neutrino is given by
\begin{equation}\label{nu_cha_rad}
{<r_{\nu}^2>}=-{6}\frac{df_{Q}(q^2)}{dq^2}{\mid_{ q^2=0}},
\end{equation}
which is determined by the second term in the expansion of the
neutrino charge form factor
\begin{equation}
f_{Q}(q^2)=f_{Q}(0)+q^2\frac{df_{Q}(q^2)}{dq^2}{\mid_{ q^2=0}}
\end{equation}
in series of powers of $q^2$.

Note that there is a long standing discussion (see \cite{Giunti:2014ixa}
for details) in the literature on the possibility to obtain
(calculate) for the neutrino charged radius a gauge independent and
finite quantity. In the corresponding calculations, performed in the
one-loop approximation including additional terms from the $\gamma-Z$
boson mixing and the box diagrams involving $W$ and $Z$ bosons, the
following gauge-invariant result for the neutrino charge radius have
been obtained \cite{Bernabeu:2004jr},
\begin{equation}{<r_{\nu_l}^2>}=\frac{G_F}{4\sqrt{2}\pi^2}\Big[3-2\log\big(\frac{m_l^2}
{m^2_W}\big)\Big],
\end{equation}
 where $m_W$ and $m_l$ are the $W$ boson and
lepton masses ($l=e,\mu,\tau$). This result, however, revived
the discussion \cite{Fujikawa:2003ww} on the
definition of the neutrino charge radius. Numerically, for the
electron neutrino electroweak radius it yields ${<r_{\nu_e}^2>}=4
\times 10^{-33} \, {cm}^2$. This theoretical result differs at most
by one order of magnitude from the available experimental bounds on
$<r_{\nu_i}^2>$ (see \cite{Giunti:2014ixa} for references and more detailed
discussion). Therefore, one may expect that the experimental
accuracy will soon reach the value needed to probe the neutrino
effective charge radius.

\section{Magnetic and electric dipole moments in gauge models}

The most well studied and understood among the neutrino electromagnetic characteristics are the dipole magnetic and electric (diagonal, $i=j$, and transition, $i\neq j$) moments
\begin{equation}
\mu_{ij}=f_{M}(0)_{ij}, \ \ \ \epsilon_{ij}=f_{E}(0)_{ij}
\end{equation}
given by the corresponding form factors  at $q^2=0$.

The diagonal magnetic and electric moments of a Dirac neutrino in the minimally-extended Standard Model with right-handed neutrinos, derived for the first time in \cite{Fujikawa:1980yx}, are respectively,
\begin{equation}\label{mu_D}
    \mu^{D}_{ii}
  = \frac{3e G_F m_{i}}{8\sqrt {2} \pi ^2}\approx 3.2\times 10^{-19}
  \Big(\frac{m_i}{1 \, \text{eV}}\Big) \mu_{B}
  \end{equation}
and
\begin{equation}\label{epsilon_D}
\epsilon^{D}_{ij}=0,
\end{equation}
where $\mu_B$ is the Borh magneton.
Note that the estimation (\ref{mu_D}) of the obtained value for the neutrino magnetic moment  shows that it is very small. It is indeed very small if compare, for instance, with the similar characteristic of electromagnetic properties of  charged leptons ($l=e, \mu, \tau $) that are the anomalous magnetic moments given by
\begin{equation}\label{AMM}
\mu^{AMM}_{l}=\frac{\alpha_{QED}}{2\pi}\mu_{B} \sim 10^{-3} \mu_{B}.
\end{equation}
There is also a reasonable gap between the prediction (\ref{mu_D}) of the minimally-extended Standard Model with right-handed neutrinos and the present experimental and astrophysical upper bounds on the neutrino effective magnetic moments. However, in many other theoretical frameworks beyond the minimally-extended Standard Model the neutrino magnetic moment can reach values being of interest for the next generation terrestrial experiments and also accessible for astrophysical observations.

The transition magnetic and electric  moments of a Dirac
neutrino are given by \cite{Shrock:1982sc,PhysRevD.25.766,Bilenky:1987ty}
\begin{equation}\label{mu_D_epsilon_D_trans}
\begin{array}{c}
    \mu^{D}_{ij}\\
    \epsilon^{D}_{ij}
  \end{array}\Bigg  \}=\frac{3e G_F m_{i}}{32\sqrt {2} \pi ^2}
  \Big(1\pm \frac{m_j}{m_i}\Big)
  \sum_{l=\ e, \ \mu, \ \tau}\Big(\frac{m_l}{m_W}\Big)^2U_{lj}U^{\ast}_{li}.
\end{equation}
Numerically  transition moments of a Dirac neutrino
can be expressed as follows
\begin{equation}
\begin{array}{c}
    \mu^D_{ij}\\
    \epsilon^D_{ij}
  \end{array}\Bigg  \}=4\times 10^{-23} \mu_{B}\Big(\frac{m_i\pm m_j}{1\ \text{eV}}\Big)
  \sum_{l=\ e, \ \mu, \ \tau}\Big(\frac{m_l}{m_\tau}\Big)^2U_{lj}U^{\ast}_{li}.
\end{equation}
The transition magnetic moment $\mu^{D}_{ij}$ is even much smaller than the diagonal magnetic moment $\mu^{D}_{ii}$ because of the leptonic GIM mechanism. That is a reason why the neutrino radiative decay $\nu_i\rightarrow \nu_j +\gamma$ is in general a very slow process.

The transition magnetic and electric moments of a Majorana neutrino are given by
\cite{Shrock:1982sc}
\begin{align}\label{mu_M_trans}
\mu^{M}_{ij}
=
-
\frac{3e G_F m_{i}}{16\sqrt{2}\pi^{2}}
\left( 1 + \frac{m_{j}}{m_{i}} \right)
\sum_{l=e,\mu,\tau}
\text{Im}\left[U^{*}_{lk} U_{lj}\right]
\,
\frac{m_{l}^{2}}{m_{W}^{2}}
\,,
\\
\label{epsilon_M_trans}
\epsilon^{M}_{ij}
=
-
\frac{3e G_F m_{i}}{16\sqrt{2}\pi^{2}}
\left( 1 - \frac{m_{j}}{m_{i}} \right)
\sum_{l=e,\mu,\tau}
\text{Re}\left[U^{*}_{lk} U_{lj}\right]
\,
\frac{m_{l}^{2}}{m_{W}^{2}}
\,.
\end{align}
There is the increase by a factor of 2 of the first coefficient
with respect to the Dirac case in (\ref{mu_D_epsilon_D_trans}). It is possible to show \cite{Schechter:1981hw, Pal:1981rm} that, depending
on the relative $CP$ phase of the two neutrinos $\nu_i$ and $\nu_j$,
one of the two options is realized:
\begin{equation}
\mu^{M}_{ij}=2\mu^{D}_{ij}, \
\epsilon^{M}_{ij}=0 \ \  \ or  \ \ \ \mu^{M}_{ij}=0, \
\epsilon^{M}_{ij}=2\epsilon^{D}_{ij}.
\end{equation}

The dependence of the diagonal
magnetic moment of a massive Dirac neutrino on the neutrino $b_{i}=m_{i}^{2}/m_{W}^{2}$
and charged lepton $a_{l}=m_{l}^{2}/m_{W}^{2}$ mass parameters in the one-loop approximation in the minimally-extended Standard Model with right-handed neutrinos was studied in  \cite{CabralRosetti:1999ad,Dvornikov:2003js,Dvornikov:2004sj}.
The calculations of the neutrino magnetic
moment which take into account exactly the dependence on the
masses of all particles \cite{Dvornikov:2003js,Dvornikov:2004sj} can be useful in the case of a heavy
neutrino with a mass comparable or even exceeding the values of
the masses of other known particles. Although the LEP data require that
the number of light neutrinos coupled to the $Z$ boson is three,
a possibility of any additional active neutrino heavier
than $m_{Z}/2$ is not excluded by current data
(see \cite{Djouadi:2012ae}).

For a heavy neutrino with mass $m_{i}$
much larger than the charged lepton masses
but smaller than the $W$-boson mass
($2 \, \text{GeV} \ll m_{i} \ll 80 \, \text{GeV}$),
 the obtained diagonal magnetic moment is \cite{Dvornikov:2003js,Dvornikov:2004sj}
\begin{equation}\label{mu_intermediate_nu}
\mu_{ii}
\simeq
\frac{3eG_{F}}{8\pi^{2}\sqrt{2}}
\,
m_{i}
\left(
1+{\frac{5}{18}}b_{i}
\right)
,
\end{equation}
whereas the result for a heavy neutrino with mass $m_{i}$
much larger than the $W$-boson mass is as follows,
\begin{equation}\label{mu_heavy_nu}
\mu_{ii}
\simeq
\frac{eG_{F}}{8\pi^{2}\sqrt{2}}
\,
m_{i}
.
\end{equation}
Note that in both cases the Dirac neutrino magnetic
moment is proportional to the neutrino mass. This is an expected
result, because the calculations have been performed within the
minimally-extended Standard Model with right-handed neutrinos.

As it has been already mentioned, the linear dependence of the neutrino magnetic moment on the neutrino mass makes its value in general very small. This feature is common for a vide class of theoretical models and seems to be hardly avoidable. There is also indeed a general problem for a theoretical model of how to get a large magnetic
moment for a neutrino and simultaneously to avoid an unacceptable large
contribution to the neutrino mass. If a contribution to the neutrino
magnetic moment of an order
\begin{equation}\label{mu_Lambda}
\mu_{\nu} \sim \frac{eG}{\Lambda},
\end{equation}
is generated by physics beyond the minimally-extended Standard Model at an
energy scale characterized by $\Lambda$, then the correspondent
contribution to the neutrino mass is \cite{Pal:1991pm,Bell:2005kz,Balantekin:2006sw}
\begin{equation}\label{mu_Lambda}
\delta m_{\nu} \sim \frac{\Lambda ^2}{2m_e}\frac{\mu_{\nu}}{\mu_B}=
\frac{\mu_{\nu}}{10^{-18}\mu_B}\Big(\frac{\Lambda}{1 \ Tev}\Big)^2\
eV.
\end{equation}
Therefore, a particular fine tuning is needed to get a large value
for the neutrino magnetic moment while keeping the neutrino mass
within experimental bounds. Different possibilities to have a large
magnetic moment for a neutrino were considered
 in the literature starting with \cite{Voloshin:1987qy,Barr:1990um}, one of the possibilities
have been considered recently \cite{Xing:2012gd}.

\section{Neutrino electromagnetic properties in reactor experiments}

The most established and sensitive
method for the experimental investigation of neutrino electromagnetic properties
is provided by direct laboratory measurements of
(anti)neutrino-electron scattering at low energies in solar,
accelerator and reactor experiments. A detailed description of
different experiments can be found in~\cite{Wong:2005pa, Balantekin:2006sw, Beda:2007hf,Giunti:2008ve, Broggini:2012df,Giunti:2014ixa}.
Here below we focus on the reactor antineutrino experiment since it provides the best nowadays terrestrial laboratory upper limit on the effective neutrino magnetic moment, as well a limit on the neutrino millicharge and probably with the expected improvement of sensitivity will be also sensitive in future to the neutrino charge radius.

In general case the cross section for an (anti)neutrino scattering on a free
electron can be written~\cite{Vogel:1989iv} as a sum of the Standard Model weak interaction and the electromagnetic interaction
contributions,
\begin{equation}
\label{cr_sec}\frac{d\sigma}{dT}=\Big(\frac{d\sigma}{dT}\Big)_{SM}+
\Big(\frac{d\sigma}{dT}\Big)_{EM}.
\end{equation}

The Standard Model contribution is
\begin{align}\label{d_sigma_SM}
\Big(\frac{d\sigma}{dT}\Big)_{SM}=\frac{G^2_F m_e}{2\pi}\Bigg[(g_V +
g_A)^2\null & \null + (g_V - g_A)^2\Big(1-\frac{T}{E_\nu}\Big)^2  \nonumber
\\ \null & \null +(g_A^2 -
g_V^2)\frac{m_eT}{E^2_\nu}\Bigg],
\end{align}
where $E_\nu$ is the initial neutrino energy and $T$ is the electron
recoil energy which is measured in the experiment. The coupling
constants are $g_V={2\sin ^2 \theta _W + \frac{1}{2}}$ and $g_A=-\frac{1}{2}$.
In the case $E_\nu\gg T$, which
is relevant to the experiments with reactor (anti)neutrinos
\begin{equation} \label{SM}
\Big(\frac{d\sigma}{dT}\Big)_{SM}= {G_F^2  m \over 2 \pi} \left( 1+ 4 \sin^2
\theta_W + 8 \sin^4 \theta_W \right) \left [ 1 + O \left ( {T
\over E_\nu} \right) \right ].
\end{equation}

The electromagnetic interaction part of the cross section can be written as a sum of three contributions originated due to the neutrino magnetic moment, millicharge and charge radius respectively,
\begin{equation}\label{d_sigma_EM}
\Big(\frac{d\sigma}{dT}\Big)_{EM}=\Big(\frac{d\sigma}{dT}\Big)_{\mu_{\nu}}+\Big(\frac{d\sigma}{dT}\Big)_{q_{\nu}}+
\Big(\frac{d\sigma}{dT}\Big)_{\langle r_{\nu_e}^2\rangle}.
\end{equation}
The contribution to the cross section generated by possible non-zero charge radius $\langle r_{\nu_e}^2\rangle$ is given by (\ref{d_sigma_SM}) with the redefined vector coupling constant \cite{Vogel:1989iv}
$g_V\rightarrow \frac{2}{3}m^2_W {\langle r_{\nu_e}^2\rangle}\sin^2 \theta_W$,
\begin{equation}\label{d_sigma_ch_rad}
\Big(\frac{d\sigma}{dT}\Big)_{\langle r_{\nu_e}^2\rangle}=\Big(\frac{d\sigma}{dT}\Big)_{{EM}_{{g_V\rightarrow \frac{2}{3}m^2_W {\langle r_{\nu_e}^2\rangle}\sin^2 \theta_W}}}.
\end{equation}
The magnetic moment contribution to the cross section and its expression at $E_\nu \gg T$ are as follows,
\begin{equation} \label{d_sigma_mu}
\Big(\frac{d\sigma}{dT}\Big)_{\mu_{\nu}}=  \pi {\alpha_{QED}^2 \over m_e^2} \left ( {\mu_\nu \over \mu_B} \right )^2
\left ( {1 \over T} - {1 \over E_\nu } \right )\approx
\pi\alpha_{QED}^{2}\frac{1}{m_{e}^{2}T}
\left(\frac{\mu_{\nu}}{\mu_{B}}\right)^{2}.
\end{equation}
The magnetic moment contribution to the cross section also changes the helicity of the neutrino, contrary to all other discussed contributions. The contribution to the cross section due to the neutrino millicharge is
\begin{equation}\label{sigma_q_e}
\left(\frac{d\sigma}{dT}\right)_{q_{\nu}}\approx 2\pi\alpha
\frac{1}{m_{e}T^2}q_{\nu}^2.
\end{equation}

Note that three terms $\Big(\frac{d\sigma}{dT}\Big)_{SM}$, \ $\Big(\frac{d\sigma}{dT}\Big)_{\mu_{\nu}}$ and $\Big(\frac{d\sigma}{dT}\Big)_{q_{\nu}}$ exhibit quite different dependence on the electron recoil energy $T$.

 The strategy of the experiment is to find an excess of events over those due to the Standard Model and other background processes. Experimental
signatures for $\mu_{\nu}$ or $q_{\nu}$ would be an excess of events between the reactor ON over OFF
samples, which exhibits an $\frac{1}{T}$ or $\frac{1}{T^2}$ energy dependence.  It is clear that the lower the measured electron recoil energy $T$ is the smaller neutrino magnetic moment $\mu_{\nu}$ and millicharge $q_{\nu}$ values can be probed in the experiment.

\begin{figure}[h]
\begin{center}
\center{\includegraphics[width=1.0\linewidth]{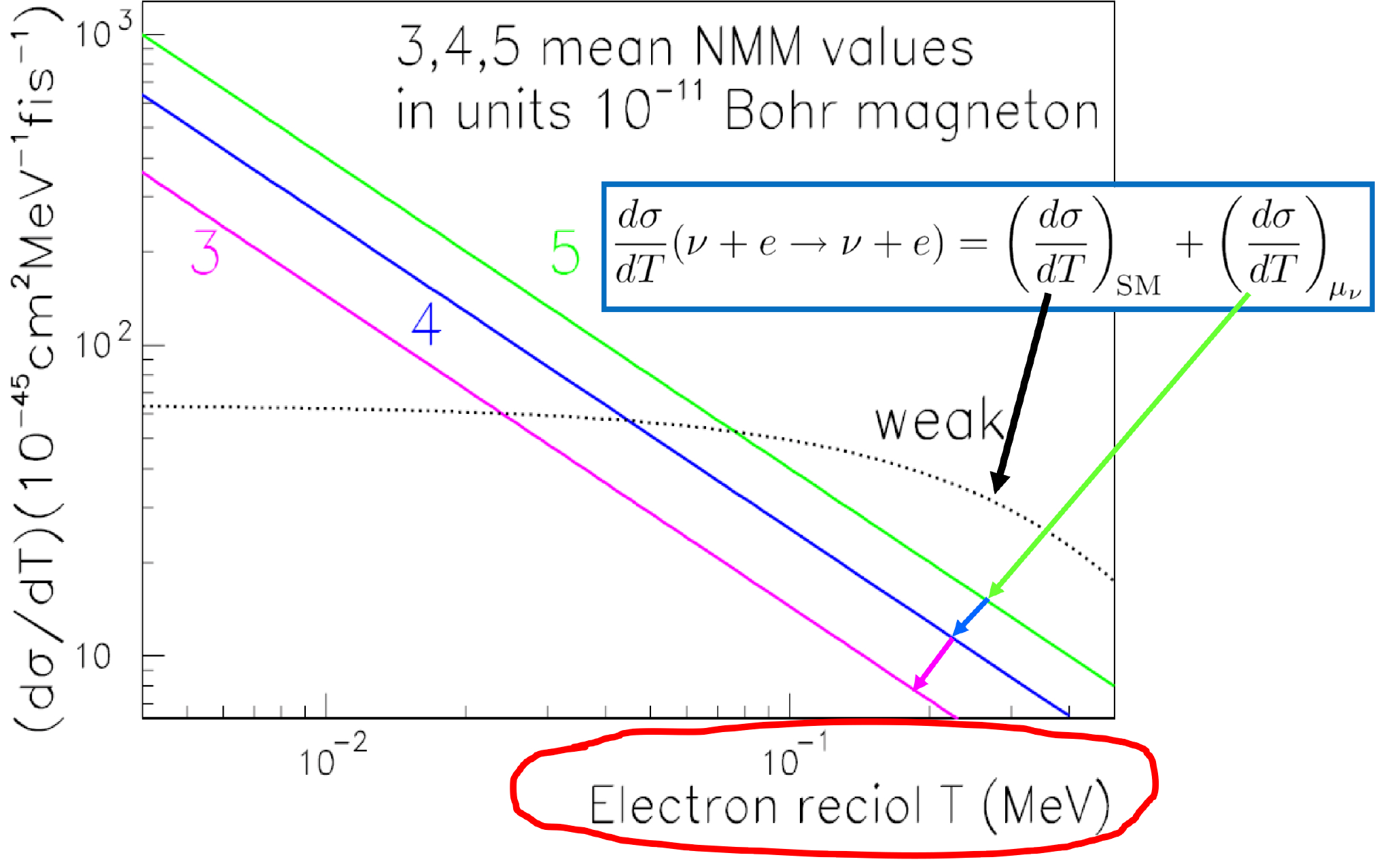}}
\caption{Standard model weak (SM) and electromagnetic magnetic moment ($\mu_\nu$) contributions to the total cross section for several values of the magnetic moment ($\mu_\nu = 3, 4, 5 \times 10^{-11} \mu_{B}$).}
\label{mu_nu_plot}
\end{center}
\end{figure}

Up to now in the reactor experiments there is no any access of events due to possible neutrino electromagnetic interactions.
From comparison of two cross sections (\ref{SM}) and (\ref{d_sigma_mu}) (the $T$ dependence of these cross sections for three fixed values $\mu_\nu = 3,4,5 \times 10^{-11}\mu_{B}$ are shown in Fig. \ref{mu_nu_plot}) the sensitivity of the experiment to the neutrino magnetic moment $\mu^2_\nu$ can be estimated:
\begin{equation}\label{mu_sensitivoty}
\mu^2_\nu\preceq \frac{G_F^2 m^3_e T}{2\pi^2\alpha_{QED}^2}(1+4\sin^2_W+8\sin^2_W)\mu_B ^2.
\end{equation}
The best laboratory upper limit on a neutrino magnetic moment \cite{Beda:2012zz} has been obtained by the GEMMA collaboration  (Germanium Experiment for measurement of Magnetic Moment of Antineutrino) that investigates the reactor antineutrino-electron scattering at the Kalinin Nuclear Power Plant (Russia). Within the presently reached electron recoil energy threshold of
\begin{equation}\label{GEMMA_T}
T \sim 2.8 \ keV
\end{equation}
the neutrino magnetic moment is bounded from above by the value
\begin{equation}\label{mu_bound}
\mu_{\nu} < 2.9 \times 10^{-11} \mu_{B} \ \ (90\% \ C.L.).
\end{equation}
This limit obtained from unobservant distortions in the recoil electron energy
spectra is valid for both Dirac and Majorana neutrinos and for both diagonal
and transitional moments.

It is planned  that with the GEMMA-II, which is now in the final stage of preparation and
is expected to get data in 2015 (for details see in \cite{Beda:2012zz,Beda:2013mta}), the effective threshold will be reduced to $T=1.5 \ keV$ and the sensitivity to the neutrino magnetic moment will be at the level
\begin{equation}\label{mu_GEMMA_II}
\mu_{\nu}\sim 1\times 10^{-11} \mu_{B}.
\end{equation}
Within further reduction of the threshold to $T= 350 \ eV$ that is now under consideration as a goal of the GEMMA-III approximately for the year 2018 the expected sensitivity of the experiment will be
at the level
\begin{equation}\label{mu_GEMMA_III}
\mu_{\nu}\sim 9\times 10^{-12} \mu_{B}.
\end{equation}

Stringent upper bounds on the neutrino magnetic moment have been also
obtained in other recently carried reactor experiments: $\mu_\nu \leq
9.0 \times 10^{-11}\mu_{B}$ (MUNU collaboration \cite{Daraktchieva:2005kn})
and $\mu_\nu \leq 7.4 \times 10^{-11}\mu_{B}$ (TEXONO collaboration
\cite{Wong:2006nx}. Stringent limits  also
obtained  in the solar neutrino scattering experiments: $\mu_\nu \leq
1.1 \times 10^{-10}\mu_{B}$ (Super-Kamiokande collaboration
\cite{Liu:2004ny}) and $\mu_\nu \leq 5.4 \times 10^{-11}\mu_{B}$
(Borexino collaboration \cite{Arpesella:2008mt}).

Interpretations and comparisons among various experiments should take into account the difference in the
flavour compositions between them at the detectors.
It should be mentioned \cite{Beacom:1999wx} that what is measured in scattering
experiments is an effective magnetic moment $\mu^{exp}_{l}$, that
depends on the flavour composition of the neutrino beam at the
detector located at a distance $L$ from the source, and which value
is a rather complicated function of the magnetic (transition) moments
$\mu_{i j}$:
\begin{equation}\label{mu_exp}\nonumber
{\mu^{exp}_{\nu_l}} ^{2} = \mu_{\nu}^{2}(\nu_{l},L,E_{\nu})=\sum_{j}\Big|
\sum_{i} U_{li} e^{-iE_{i}L}(\mu_{ji}-i\epsilon_{ji}) \Big| ^{2}.
\end{equation}
The dipole electric (transition) moments, if these quantities not
vanish, can also contribute to $\mu^{exp}_{l}$. The detailed discussion of this item is given in
\cite {Giunti:2014ixa}.

The absence of distortions of the electron recoil energy spectra in GEMMA experiment can be also used  \cite{Studenikin:2013my} to bound from above the neutrino millicharge $q_{\nu}$. From  demanding that possible effect due to $q_{\nu}$ does not exceed one due to $\mu_{\nu}$ and comparing two cross sections given by (\ref{d_sigma_mu}) and (\ref{sigma_q_e}) for the sensitivity of the experiment to $q_{\nu}$ we get
\begin{equation}\label{q_limit}
q_{\nu}^{2}\preceq\frac{T}{2m_e}\left(\frac{\mu^{a}_{\nu}}{\mu_{B}}\right)^{2}e_0 ^2.
\end{equation}
From (\ref{q_limit}) and the present GEMMA data given by (\ref{mu_bound}) and (\ref{GEMMA_T}) a bound is derived  \cite{Studenikin:2013my},
\begin{equation}
\mid q_{\nu}\mid \sim  1.5 \times 10^{-12} e_0.
\end{equation}
Note that this bound should be considered as an order-of-magnitude estimation for a possible sensitivity of the experiment to $q_{\nu}$. More accurate analysis accounting  for experimental data taken over an extended energy range using the corresponding statistical procedures give the following limit \cite{Studenikin:2013my}
\begin{equation}
\mid q_{\nu} \mid < 2.7 \times 10^{-12} e_0 \ (90 \% \ C.L.).
\end{equation}
More stringent constraints on the millicharge on the level of $q_{\nu}\sim \ few \ \times 10^{-13}e_{0}$  are expected with further progress in antineutrino-electron cross section measurements with GEMMA-II and III.
Finally, note that upper bounds on the neutrino electric millicharge on the level of $\mid q_{\nu} \mid \sim 10^{-12} e_0$ are also discussed in \cite{Chen:2014dsa}.



\section{Neutrino electromagnetic interactions in astrophysics}

If a neutrino has
the non-trivial electromagnetic properties discussed above, a direct
neutrino coupling to photons is possible and several processes important
for applications in astrophysics exist \cite{Raffelt:1996wa}. A set of
most important neutrino electromagnetic processes is: 1) neutrino
radiative decay $\nu_{1}\rightarrow \nu_{2} +\gamma$, neutrino
Cherenkov radiation in an external environment (plasma and/or
electromagnetic fields), spin light of neutrino, $SL\nu$ , in the
presence of a medium \cite{Lobanov:2002ur,Studenikin:2004dx,Grigorev:2005sw,Grigoriev:2012pw}; \ 2) photon (plasmon) decay to a
neutrino-antineutrino pair in plasma $\gamma \rightarrow \nu {\bar
\nu }$; \ 3) neutrino scattering off electrons (or nuclei); \ 4)
neutrino spin (spin-flavor) precession in a magnetic field (see
\cite{Okun:1986na}) and resonant neutrino spin-flavour
oscillations in matter \cite{Lim:1987tk,Akhmedov:1988uk}.

The tightest astrophysical bound on a neutrino magnetic moment is
provided by observed properties of globular cluster stars. For a
large enough neutrino magnetic moment the plasmon decay rate can be
enhanced so that a reasonable delay of helium ignition would appear.
From lack observation evidence of anomalous stellar cooling due to
the plasmon decay the following limit has been found \cite{Raffelt:1990pj}
\begin{equation}
\Big( \sum _{i,j}\mid \mu_{ij}\mid ^2\Big) ^{1/2}\leq 3 \times
10^{-12} \mu _B.
\end{equation}
This is the most stringent astrophysical constraint on neutrino
magnetic moments, applicable to both the Dirac and Majorana neutrinos.

A new interesting phenomena of neutrino electromagnetic interactions in astrophysical environments has been recently considered \cite{Studenikin:2012vi}. It is shown that millicharged neutrinos move on curved trajectories inside dense magnetized rotating stars. The feedback of the neutrino flux should effect the rotation of pulsars. This phenomena has been termed the ``Neutrino Star Turning'' ($\nu ST$) mechanism~\cite{Studenikin:2012vi}. A new limit on the neutrino millicharge
\begin{equation}
q_0<1.3\times10^{-19}e_0
\end{equation}
is obtained in order to avoid a contradiction of the $\nu ST$ impact with observational data on pulsars. This limit is among the strongest astrophysical constraints on the neutrino millicharge.

\section{Acknowledgments}
One of the authors (A.S.) is thankful to Arcadi Santamar\'{\i}a, Salvador Mart\'{\i} and Juan Fuster for the kind invitation to participate at the ICHEP 2014  conference and to all of the organizers for their hospitality in Valencia. This study has been partially supported by the Russian Foundation for Basic Research (grant No. 14-22-03043-ofi).




\bibliographystyle{model1a-num-names}
\bibliography{Stu_NPB_PS}







\end{document}